# Superluminal Tunneling of a Relativistic Half-Integer Spin Particle Through a Potential Barrier

Luca Nanni*

*Correspondence: luca.nanni@student.unife.it

**Abstract**: This paper investigates the problem of a relativistic Dirac half-integer spin free particle tunneling through a rectangular quantum-mechanical barrier. If the energy difference between the barrier and the particle is positive, and the barrier width is large enough, there is proof that the tunneling may be superluminal. For first spinor components of particle and antiparticle states, the tunneling is always superluminal regardless the barrier width. Conversely, the second spinor components of particle and antiparticle states may be either subluminal or superluminal depending on the barrier width. These results derive from studying the tunneling time in terms of phase time. For the first spinor components of particle and antiparticle states, it is always negative while for the second spinor components of particle and antiparticle states, it is always positive, whatever the height and width of the barrier. In total, the tunneling time always remains positive for particle states while it becomes negative for antiparticle ones. Furthermore, the phase time tends to zero, increasing the potential barrier both for particle and antiparticle states. This agrees with the interpretation of quantum tunneling that the Heisenberg uncertainty principle provides. This study's results are innovative with respect to those available in the literature. Moreover, they show that the superluminal behaviour of particles occurs in those processes with high-energy confinement.

## 1   Introduction

Several theoretical and experimental studies in the past decades have examined phenomena involving superluminal waves and objects because of their implication in quantum and cosmological physics [1-6]. Among them, the study of the tunneling time problem is one of the topics that has most attracted the interest of quantum physicists [7-12]. Researchers have approached this issue both from the perspective of non-relativistic [13] and relativistic [14] quantum theory. In both cases the tunneling time does not depend on the barrier width (at least for large enough barriers), thus proving superluminal behaviour of the quantum object (wave or particle). However, the tunneling time problem remains a controversial one in quantum physics. A comprehensive and clear theory to explain *how long does it take a particle to tunnel through a barrier* still does not exist [15]. As is well known, *classical* quantum mechanics does not treat time as an Hermitian operator but rather as a parameter [16]. Time does not appear in the commutation relationships typical of the Hermitian operators, even if it appears in one of the forms of the Heisenberg uncertainty principle, being a physical variable conjugated to the energy. For this reason we have to give up directly knowing the tunneling time. We may bypass the obstacle by assuming that the wave packet inside the barrier is stationary, with an imaginary wave vector. We can then interpret the tunneling time as the phase variation of the evanescent stationary wave that crossing the potential barrier induces. This is the definition of tunneling phase time [17-18] and is the most widely used quantity in the studies of tunneling phenomena.

In this paper we investigate, by the study of the phase time, the one-dimensional scattering process of a relativistic half-integer spin free particle through a rectangular barrier. Because the motion takes place in one dimension, particle and antiparticle states are bi-spinor (there is no spin-flip in one-dimensional motion). Thus, we investigate the tunneling time for each of the spinor components. When the barrier height is greater than the particle energy, for the first component of particle and antiparticle states, we prove that the tunneling process always occurs at superluminal velocity (negative tunneling times). This transpires regardless of the width of the barrier and the energy gap between the barrier and the relativistic particle. Second spinor components of particle and antiparticle states, on the contrary, behave in a different way. In these, the scattering through the barrier may be subluminal or superluminal depending on the barrier width (positive tunneling time). In total, the tunneling time is almost always positive for particle states, while it becomes negative for antiparticle ones. Therefore, the Hartman effect always occurs for both particles and antiparticles. Furthermore, the study shows that when the energy gap increases, the tunneling time tends to zero for both particle and antiparticle states, and the crossing velocity of the barrier diverges to infinite values. Finally, we prove that superluminality is more accentuated for antiparticle states than particle states. This peculiarity disappears as the particle velocity increases, i.e., when the second spin component becomes more and more important (its magnitude increases as the relativistic factor $\beta = u/c$ increases).

## 2 Quantum Tunneling from the Perspective of the Heisenberg Uncertainty Principle

In 1928 George Gamow gave a simple but elegant interpretation of quantum tunneling [18]: the particle with energy $E$ *borrows* from the *vacuum* an amount of energy $(U - E)$ to surmount the barrier potential $U > E$. Recalling the uncertainty principle in the form:

$$\partial E \partial t \geq \hbar \tag{1}$$

provides the time needed to *repay* the *borrowed* energy as:

$$\partial t = \frac{\hbar}{(U-E)} \tag{2}$$

Eq. **(2)** gives the tunneling time, i.e., the time the quantum particle needs to cross the potential barrier. It is interesting to note that this time does not depend on the geometry of the barrier, providing simple evidence of the Hartman effect [11]. If $a$ is the one-dimensional barrier width, the tunneling velocity is:

$$v_t = \frac{a}{\partial t} = \frac{a(U-E)}{\hbar} \tag{3}$$

Since the barrier width and the energy gap can assume whatever value, from Eq. **(3)** we conclude that there are geometrical and energetic conditions under which the tunneling is a superluminal process. In particular, for large values of $a$ (high confinement) and $(U - E)$ (high-energy systems), the tunneling velocity inside the barrier is higher than the speed of light, according to a previous work by the present author [20].

## 3 The Scattering Model

Let us consider a half-integer spin relativistic free particle of rest mass $m_0$, energy $\pm E$ (depending on its particle or antiparticle state) and velocity $u$, scattering through a potential barrier of width $a$ and potential energy $U$. The potential energy is always greater than the particle energy ($U > \pm E$):

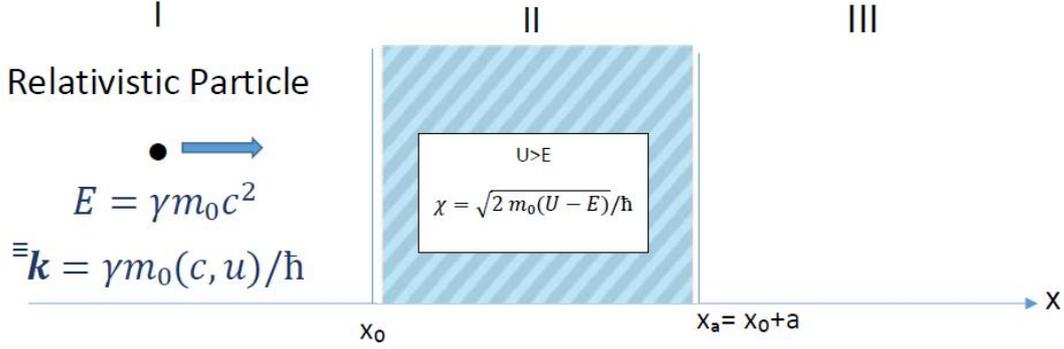

**Figure 1**: Particle scattering through a potential barrier: in regions I and III the particle behaves as a free relativistic quantum object. In region II the particle behaves in a transcendent way.

In regions I and II, where the potential energy is zero, the particle moves freely and behaves like a relativistic monochromatic de Broglie wave plane [21], which spreads in space-time. To simplify the study we suppose a two-dimensional space-time, so that the plane wavefront impinging on the barrier has the two-component spinor form (there is no spin-flip):

$$\begin{cases} |\Psi\rangle_+ = N \begin{pmatrix} 1 \\ \frac{cp_x}{E+m_0c^2} \end{pmatrix} \exp\{i(k_0ct - k_xx)\} = N \begin{pmatrix} 1 \\ \beta \end{pmatrix} \exp\{i(k_0ct - k_xx)\} \\ |\Psi\rangle_- = N \begin{pmatrix} -\frac{cp_x}{E+m_0c^2} \\ 1 \end{pmatrix} \exp\{-i(k_0ct - k_xx)\} = N \begin{pmatrix} -\beta \\ 1 \end{pmatrix} \exp\{-i(k_0ct - k_xx)\} \end{cases} \quad (4)$$

In Eq. **(4)**, $\beta$ is the relativistic factor ($u/c$), while $|\Psi\rangle_+$ and $|\Psi\rangle_-$ are respectively the particle and antiparticle spinors. $N$ is the normalization constant that, for a given finite volume, $\sqrt{(\gamma+1)/2\gamma V}$ provides, where $\gamma$ is the Lorentz factor. Wave vectors $k_0$ and $k_x$ are:

$$\begin{cases} k_0 = \frac{p_0}{\hbar} = \frac{\gamma m_0 c}{\hbar} \\ k_x = \frac{p_x}{\hbar} = \frac{\gamma m_0 u}{\hbar} \end{cases} \quad (5)$$

At the beginning of the barrier ($x = x_0$), the de Broglie wave may be reflected or transmitted. In the latter case, the wave behaves like an evanescent stationary wave [13]. Since the wave is confined within the barrier (evanescent waves are not propagating waves), its wave vector has only a spatial component:

$$\begin{cases} \chi_+ = \frac{\sqrt{2m_0(U-E)}}{\hbar} \quad Particle \\ \chi_- = \frac{\sqrt{2m_0(U+E)}}{\hbar} \quad Antiparticle \end{cases} \quad (6)$$

At the end of the barrier, the evanescent wave is transmitted in region III, where it returns to be a free relativistic particle with half-integer spin. Since the wave function **(4)** is a two-component vector, we must study the scattering process for both of them. At the barrier edges, each component of the Dirac spinor has to match the continuity conditions that quantum theory imposes.

We will investigate the tunneling time by the phase time (which is asymptotic and non-local) defined as [11]:

$$\tau = \hbar \frac{\partial arg\{C_T exp[i(k_0 ct_a - k_x x_a)]\}}{\partial E} \quad (7)$$

Therefore, the study object of this research preliminarily requires the calculation of the transmission amplitude $C_T$ for the spinor component of the particle and antiparticle states.

## 4 Transmission and Reflection Amplitudes

Let us calculate the transmission and reflection amplitudes of the scattering process that we will use for tunneling time investigation. Considering initially the first component of the particle spinor **(4)**, the continuity conditions are:

$$\begin{cases} \Psi(I)|_{x_0} = \varphi(II)|_{x_0} \\ \Psi(III)|_{x_0+a} = \varphi(II)|_{x_0+a} \\ \Psi'(I)|_{x_0} = \varphi'(II)|_{x_0} \\ \Psi'(III)|_{x_0+a} = \varphi'(II)|_{x_0+a} \end{cases} \quad (8)$$

where:

$$\begin{cases} \Psi(I) = exp\{i(k_0 ct - k_x x)\} + c_R exp\{-i(k_0 ct - k_x x)\} \\ \varphi(II) = \alpha e^{\chi x} + \delta e^{-\chi x} \\ \Psi(III) = c_T exp\{i(k_0 ct - k_x x)\} \end{cases} \quad (9)$$

$c_R$ and $c_T$ are respectively the reflection and the transmission coefficients of the particle wave, while $\alpha$ and $\delta$ are the coefficients of the evanescent wave (see Appendix A). Substituting functions **(9)** in **(8)** and solving the system by the Cramer algorithm, we get:

$$c_T = \frac{4i\chi\Delta k}{exp\{i(k_0 ct_a - k_x x_a)\}[e^{\chi x_a}(\chi - i\Delta k)^2 + e^{-\chi x_a}(\Delta k - i\chi)^2]} \quad (10)$$

$$C_R = \frac{exp\{2i(k_0 ct_0 - k_x x_0)\}exp\{i(k_0 ct_a - k_x x_a)\}}{4(\chi^2 + \Delta k^2)sinh^2(\chi x_a)}[-2sinh(\chi x_a)(\chi^2 - \Delta k^2) - 4iexp\{i(k_0 ct_0 - k_x x_0)\}exp\{i(k_0 ct_a - k_x x_a)\}cosh(\chi x_a)] \quad (11)$$

where $\Delta k = (k_0 - k_1) = \gamma m_0 (c - u)/\hbar$. For antiparticle states, Eq. **(10)** provides the transmission amplitude, replacing $\Delta k$ with $-\Delta k$ and using the evanescent wave vector $\chi_-$. Through some algebraic manipulations, we use the explicit forms **(5)** and **(6)** of the Dirac wave vectors and set $m_0 \equiv c \equiv \hbar = 1$. Eq. **(10)** thus provides the probability transmission of the particle:

$$C_T^2 = \frac{\gamma^2(1-u)^2(U-E)}{[\gamma^2(1-u)^2 + 2(U-E)]sinh^2(x_a\sqrt{2(U-E)})} \quad (12)$$

When increasing the barrier height, i.e., increasing the potential energy $U$, the probability that the particle is transmitted through the barrier rapidly goes to zero, while if $U = E$ this probability tends to one:

$$\begin{cases} \lim_{U \to \infty} C_T^2 = 0 \\ \lim_{U \to E} C_T^2 = 1 \end{cases} \quad (13)$$

The limits in **(13)** also hold for the antiparticle state, but for $U \to \infty$ the probability tends to zero more rapidly. The reflection probability $C_R^2$ has an opposite trend to the transmitted one, which the quantum relation $(C_T^2 + C_R^2) = 1$ has to satisfy.

## 5 Tunneling Time for Particle States

As mentioned earlier, the phase time **(7)** helps perform the investigation of the tunneling time. The first step for its calculation is to determine the argument of the complex number $C_T exp[i(k_0 ct_a - k_x x_a)]$. To simplify the formalism, from now on the Greek letter $\theta$ will represent this. Through some algebraic calculations (see Appendix B), we get:

$$\theta = arctg\left[-2\chi\Delta k \frac{1}{\Delta k^2 + \chi^2 \frac{e^{2\chi x_a}-1}{e^{2\chi x_a}+1}}\right] \qquad (14)$$

Assuming that the barrier width is large enough, we may set the term $(e^{2\chi x_a} - 1/e^{2\chi x_a} + 1)$ equal to one, so that the argument function becomes:

$$\theta = arctg\left[-\frac{2\chi\Delta k}{\Delta k^2 + \chi^2}\right] \qquad (15)$$

Performing its derivative respect for the particle energy and substituting the result in Eq. **(7)**, we get the phase time (see Appendix C):

$$\tau^{(1°)} = \frac{2m_0 \Delta k E(1-2\chi^2) + 2\hbar^2 \chi^2 \Delta k(2\Delta k^2 - 1)}{\hbar \chi E[(\Delta k^2 + \chi^2)^2 + 4\Delta k^2 \chi^2]} \qquad (16)$$

where the superscript $(1°)$ means that the phase time refers the first component of the Dirac spinor. Eq. **(16)** is independent from the barrier width $a$, and as expected, the Hartman effect [11] occurs. Moreover, the value of the phase time is always negative whatever the particle energy and the barrier height. According to references [22-23], this means the tunneling always occurs at superluminal velocities. **Figure 2** shows the phase time trends for $U = 10$ and $U = 100$ vs the relativistic factor $\beta$ (as usual we set $m_0 \equiv c \equiv \hbar = 1$):

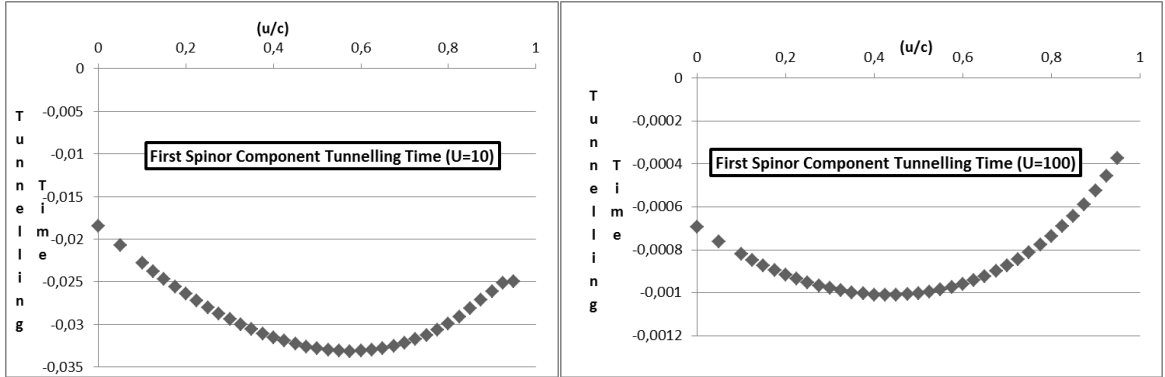

**Figure 2**: First spinor component tunneling time for particle state vs relativistic factor $\beta = (u/c)$ for two potential energies. All values are negative, proving that the tunneling is always superluminal. As the barrier height increases, the tunneling time tends to zero.

The trends show a minimum tunneling time that shifts towards low values of the relativistic factor as the barrier height increases. Moreover, when increasing the barrier height, the phase time goes rapidly to zero. This result is in agreement with the superluminality that the Heisenberg uncertainty principle predicts. The ratio between the barrier width and the tunneling time provides the barrier-crossing speed, at least in principle. From the obtained results, we thus conclude that the higher the barrier is, the faster the scattering through it is. This phenomenon occurs as if the potential barrier *boosts* the particle towards higher velocities, even if its occurrence probability (that the squared transmission amplitude provides) goes to zero as the barrier height decreases.

We may also display the superluminal tunneling of the first spinor component by the refraction index, calculated as the ratio between the velocity of the free particle before impinging on the barrier and its velocity during the tunneling:

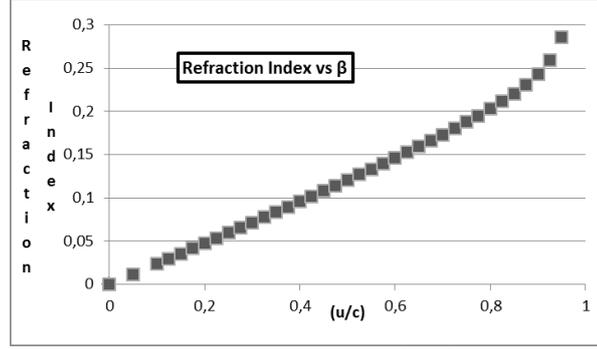

**Figure 3**: Tunneling refraction index vs relativistic factor $\beta = (u/c)$ for the barrier height $U = 10$. All values are lower than one, proving that the tunneling is always superluminal.

As expected, the refraction index is always lower than one and increases with the relativistic factor. This means that as the velocity of the free particle approaches the speed of light, the superluminality of the tunneling tends to increase, and the particle velocity inside the barrier becomes infinite. This result is explainable by the fact that by keeping the potential barrier constant and increasing the particle velocity, the energy gap rapidly tends to zero.

Let us calculate the phase time for the second component of the spinor **(4)** for the particle state following the same physico-mathematical approach used for the first component (see Appendix D):

$$\tau^{(2°)} = \hbar \frac{2(1-\beta^2)(\Delta k^2 + \chi^2)}{E[(\Delta k^2 + \chi^2)^2 + 4\beta^4 \Delta k^2 \chi^2]} + \tau^{(1°)} \tag{17}$$

This shows that the two components of the spinor *emerge* from the far side of the barrier with a phase time difference given by:

$$\Delta\tau = \hbar \frac{2(1-\beta^2)(\Delta k^2 + \chi^2)}{E[(\Delta k^2 + \chi^2)^2 + 4\beta^4 \Delta k^2 \chi^2]} \tag{18}$$

This quantity, which from now we will call phase time delay, is always positive. **Figure 4** shows the trend of the second spinor component for the particle state:

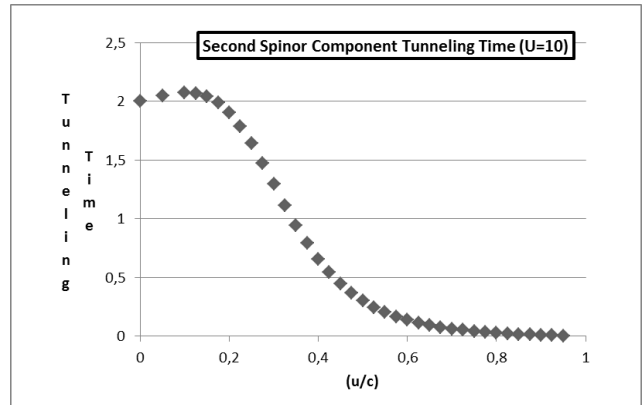

**Figure 4**: Second spinor component tunneling for the particle state for the barrier height $U = 10$. All values are positive, proving that the tunneling may be superluminal depending on the barrier width.

This delay is considerable at non-relativistic regimes ($u \ll c$), at which the magnitude of the secondary component of spinor **(4)** is very low, and tends to zero when the particle velocity approaches the speed of light. For this reason, whatever the barrier width, the secondary spinor component always *emerges* later than the principal one. This phenomenon leads to an anomalous *distortion* of the Dirac wave function that disappears when the relativistic behaviour of the particle becomes relevant. Overall, the phase time **(17)** is always positive, except when the particle velocity approaches the speed of light, and follows the trend shown in **Figure 5**:

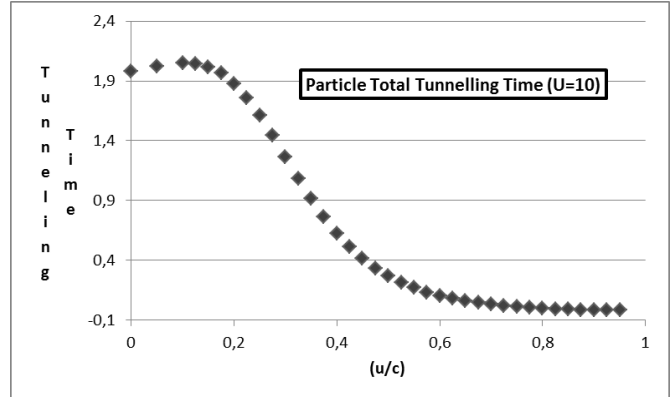

**Figure 5**: Total tunneling time from the contribution of the particle spinor components ($U = 10$). The time is almost always positive and asymptotic.

As expected, the total tunneling time is asymptotic, and the Hartman effect may occur depending on the barrier width and on the particle energy.

## 6 Tunneling Time for Antiparticle States

Let us calculate the tunneling time for the antiparticle states following the same approach used in the previous section. The negative values of energy $E$ and the relativistic wave number $\Delta k$ represent the only difference. Eq. **(16)** can then provide the phase time by replacing $E$ with $-E$, $\Delta k$ with $-\Delta k$ and using the second variant of Eq. **(6)**:

$$\tau_-^{(1°)} = -\frac{2m_0 \Delta k E(2\chi^2-1)+2\hbar^2\chi^2\Delta k(1-2\Delta k^2)}{\hbar\chi E[(\Delta k^2+\chi^2)^2+4\Delta k^2\chi^2]} \tag{19}$$

The negative sign of $\Delta k$ is due to the fact that the sign of the argument of the exponential function for the antiparticle state is opposite to that of the particle state. Eq. **(19)** does not depend on the barrier width, and the Hartman effect occurs. **Figure 6** shows the trend for $U = 10$:

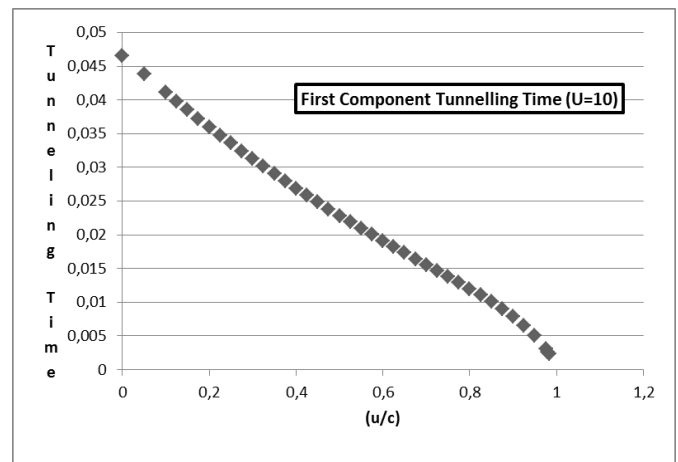

**Figure 6**: Phase time vs relativistic factor $\beta$ for the antiparticle state for the barrier height $U = 10$. The phase time is always positive and tends to zero as the antiparticle velocity approaches the speed of light.

Unlike the first spinor component of the particle state, the phase time is always positive, and the tunneling through the barrier could be subluminal or superluminal depending on the barrier width. This is expected behaviour considering that the first spinor component of the antiparticle is the opposite of the second spinor component of the particle. After setting the barrier height, the phase time tends to zero as the velocity approaches the speed of light, just like what occurs for the particle state.

Similarly, Eq. **(18)** provides the phase time difference between the two components of the antiparticle spinor, replacing $E$ with $-E$ and using the second part of Eq. **(6)**. **Figure 7** shows its trend:

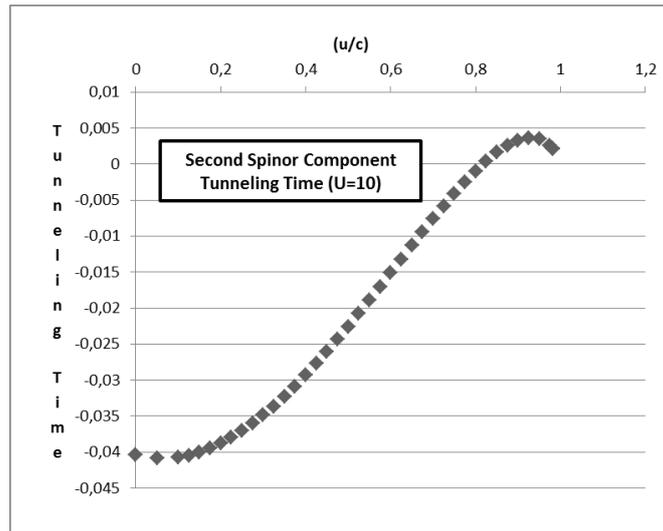

**Figure 7**: Phase time difference of the two antiparticle spinor components for the barrier height $U = 10$. The quantum tunneling effect twists the spinor, and the deformation disappears as the free particle velocity increases.

Once again unexpectedly, the secondary component of the antiparticle spinor *emerges* from the far side of the barrier before the principal one. This always leads to an anomalous *distortion* of the Dirac antiparticle spinor that, however, is opposite to the one that occurs for the particle state. This *distortion* disappears as the antiparticle velocity approaches the speed of light but is slower than what occurs for the particle spinor.

In total, the antiparticle phase time given by the contribution of the two spinor components is always negative and turns out to be positive only when the particle velocity approaches the speed of light. **Figure 8** shows the asymptotic trend:

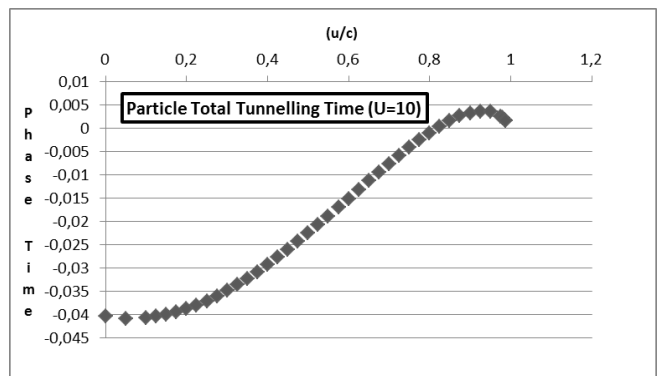

**Figure 8**: Total tunneling time from the contribution of the antiparticle spinor components ($U = 10$). The time is almost always negative and asymptotic.

## 7 Discussion

This study proves that relativistic free particles and antiparticles behave differently when a potential barrier scatters them, even at velocities close to the speed of light. For large barrier widths, these behaviours tend to converge and become uniform. Although the Hartman effect occurs for both particles and antiparticles, the tunneling is always superluminal for antiparticle states (negative phase time) except when the particle velocity approaches the speed of light (positive phase time tending to zero). However, it can be subluminal or superluminal for particle states depending on the barrier width, as the phase time is always positive except when the particle velocity approaches the speed of light (negative phase time tending to zero). The interaction of the relativistic wave packet with the potential barrier generates superluminality. Inside this, forward and backward evanescent waves describe the particle or antiparticle. As references [13, 24] suggest, this interaction leads to a complex set of interferences between the incident and reflected waves that cause an *acceleration* of the incoming wave. However, the Heisenberg uncertainty principle may also explain superluminal tunneling. The energy confinement of the particle within the barrier reduces the time uncertainty, and consequently increases the one affecting the energy. Such high uncertainty leads to the transmitted wave exceeding the speed of light. We think this interpretation of the phenomenon we are investigating is more understandable and closer to quantum theory.

We must still explain the reason why an antiparticle behaves differently than a particle, i.e., why it almost always tunnels in superluminal mode. The negative energy of the antiparticle decreases the transmission probability that Eq. **(12)** provides. This is especially due to the strong increases of the hyperbolic function at the denominator and changes the interference mechanism between forward and backward waves. In other words, the barrier tends to reflect an antiparticle more than a particle. An antiparticle tunneling through the potential barrier $U$ is equivalent to a particle that tunnels through a potential barrier $(U + E)$. This provides a more physical explanation of why an antiparticle has a lower probability of penetrating the barrier. However, once the antiparticle overcomes the barrier edge, it becomes more energy *confined* than the particle and *accelerates* more strongly. In fact, comparing the trend of **Figure 5** with that of **Figure 8** shows that the tunneling time of the antiparticle state is much closer to zero than the particle one, whatever its energy is. This means the superluminality of the antiparticle state is greater than the superluminality of the particle state.

If the high-energy confinement (very high barriers) facilitates acceleration towards superluminal velocity, why does increasing $\beta$ tend to decrease the phase time? Increasing the relativistic factor decreases the energy gap between the barrier and the particle/antiparticle. This reduces the energy *confinement* and, consequently, the uncertainty affecting the energy $E$. But because $E$ is so high, it is small enough of an uncertainty to *boost* the particle towards ever higher superluminal velocities. In fact, the trends show that accelerating the particle/antiparticle within the barrier is a more important kinetic factor than the energy *confinement*.

Finally, the different mechanism of interference between the transmitted and reflected waves may also explain spinor *distortion*. This provides different *accelerations* for the principal and secondary spinor components [24]. We can also state that the potential barrier contributes to spreading the wave packet. An earlier occurrence or delay of the secondary spinor component is equivalent to frequency changes, i.e., energy changes between the plane waves forming the wave packet.

## 8   Conclusion

Summarizing, we have considered tunneling of a relativistic free particle with half-integer spin through a one-dimensional potential barrier. The analysis suggests that the total transmission is almost always superluminal for antiparticle states and may be subluminal or superluminal for particle states. Within the potential barrier, the two components of the spinor behave as evanescent modes, and the high energy *confinement* accelerates the particle/antiparticle further towards the speed of light. Considering the interference between forward and backward evanescent waves may explain the phenomenon, but an easier explanation holds if we account for the Heisenberg uncertainty principle. As the particle/antiparticle velocity approaches the speed of light, the tunneling time tends to zero, making the acceleration divergent towards infinite values. The study also proves that the two spinor-transmitted components *emerge* from the potential barrier at different times, providing an anomalous *distortion* of the Dirac wave function. This *distortion* occurs in an opposite way for particles and antiparticles. More precisely, the secondary spinor component is delayed compared with the first one for particle states and is anticipated for antiparticle states. All of these results are innovative with respect to those available in the literature. They suggest a way to obtain a source of superluminal massive elementary particles: ultra-relativistic particles scattering through huge potential barriers. As discussed above, the probability of total transmission of the particle/antiparticle through the barrier strongly decreases as its incoming velocity approaches the speed of light. However, we cannot exclude the occurrence of the model used in this study a priori in a cosmological scenario.

**Appendix A**

The Cramer rule can solve linear system **(8)**. To simplify the mathematical formalism, we set:

$$\begin{cases} l_1 = exp\{i(k_0 c t_0 - k_x x_0)\} \\ l_2 = exp\{-i(k_0 c t_0 - k_x x_0)\} \\ m_1 = e^{\chi x_0} \\ m_2 = e^{-\chi x_0} \\ n_1 = e^{\chi x_a} \\ n_2 = e^{-\chi x_a} \\ q_1 = exp\{i(k_0 c t_a - k_x a_0)\} \end{cases} \quad \textbf{(A1)}$$

The determinant of the coefficient matrix $A$ is:

$$det(A) = l_2 m_2 n_1 q_1 (\chi - i\Delta k)^2 + l_2 m_1 n_2 q_1 (i\Delta k - \chi)^2 \quad \textbf{(A2)}$$

The four unknown quantities, i.e., the coefficients of the linear combination are:

$$\begin{cases} C_R = \frac{l_1 q_1}{det(A)} [(n_1 m_2 - m_1 n_2)(\chi^2 - \Delta k^2) - 2 i q_1 \Delta k \chi (m_1 n_2 + m_2 n_1)] \\ C_T = -4i \frac{l_1 l_2 n_1 n_2 \Delta k \chi}{det(A)} \\ \alpha = 2 \frac{[l_1 l_2 n_2 q_1 \Delta k (\Delta k - i\chi)]}{det(A)} \\ \delta = -2 \frac{[l_1 l_2 n_1 q_1 \Delta k (\Delta k + i\chi)]}{det(A)} \end{cases} \quad \textbf{(A3)}$$

**Appendix B**

The calculation of phase time requires knowledge of the complex number argument:

$$z = C_T exp[i(k_0 c t_a - k_x x_a)] \quad \textbf{(B1)}$$

Using **(A1)** and considering that $(\chi - i\Delta k)^2 / i = (i\Delta k - \chi)^2$ and $(\Delta k - i\chi)^2 / i = (i\chi - \Delta k)^2$, we get:

$$z = -\chi \Delta k \frac{1}{e^{\chi x_a}(i\Delta k - \chi)^2 + e^{-\chi x_a}(i\chi - \Delta k)^2} \quad \textbf{(B2)}$$

Solving the squares of the two binomials in **(B2)**, we obtain:

$$z = -\chi \Delta k \frac{1}{\Delta k^2 (e^{\chi x_a} + e^{-\chi x_a}) + \chi^2 (e^{\chi x_a} - e^{-\chi x_a}) - 2i\chi \Delta k (e^{\chi x_a} + e^{-\chi x_a})} \quad \textbf{(B3)}$$

Denoting by $z'$ the denominator of **(B3)** and recalling that its inverse is $z'^{-1} = \bar{z'}/|z'|^2$ and that $arctg(\beta/\alpha)$ provides the argument of a complex number $z = \alpha + i\beta$, we obtain:

$$arg(z) = arctg \left[ -2\chi \Delta k \frac{(e^{\chi x_a} + e^{-\chi x_a})}{\Delta k^2 (e^{\chi x_a} + e^{-\chi x_a}) + \chi^2 (e^{\chi x_a} - e^{-\chi x_a})} \right] \quad \textbf{(B4)}$$

Some simple algebraic manipulations allow us to rewrite **(B4)** as:

$$arg(z) = arctg \left[ -2\chi \Delta k \frac{1}{\Delta k^2 + \chi^2 \frac{e^{2\chi x_a} - 1}{e^{2\chi x_a} + 1}} \right] \quad \textbf{(B5)}$$

If the barrier width $a$ is large enough, then the fraction $\frac{e^{2\chi x_a}-1}{e^{2\chi x_a}+1} \cong 1$, and **(B5)** becomes:

$$arg(z) = arctg\left[-2\frac{\chi\Delta k}{\Delta k^2+\chi^2}\right] \tag{B6}$$

which is independent of the barrier width.

## Appendix C

In this section we explain the steps to get the derivative of the argument function **(B6)**. Using the chain rule shows:

$$\frac{\partial arg(z)}{\partial E} = \frac{1}{1+\frac{4\Delta k^2\chi^2}{(\Delta k^2+\chi^2)^2}}\frac{\partial}{\partial E}\left(-2\frac{\chi\Delta k}{\Delta k^2+\chi^2}\right) = \frac{(\Delta k^2+\chi^2)^2}{(\Delta k^2+\chi^2)^2+4\Delta k^2\chi^2}\frac{\partial}{\partial E}\left(-2\frac{\chi\Delta k}{\Delta k^2+\chi^2}\right)$$
**(C1)**

The last derivative of **(C1)** is:

$$\frac{\partial}{\partial E}\left(-2\frac{\chi\Delta k}{\Delta k^2+\chi^2}\right) = \frac{-2(\chi'\Delta k+\chi\Delta k')(\Delta k^2+\chi^2)+2\chi\Delta k(2\Delta k\Delta k'+2\chi\chi')}{(\Delta k^2+\chi^2)^2} \tag{C2}$$

Now we calculate the derivatives $\chi'$ and $\Delta k'$ separately using the first parts of Eq. **(5)** and **(6)** and by performing some simple algebraic manipulations:

$$\begin{cases} \chi' = \frac{\partial}{\partial E}\frac{\sqrt{2m_0(U-E)}}{\hbar} = -\frac{m_0}{\hbar^2\chi} \\ \Delta k' = \frac{\partial}{\partial E}\frac{\gamma m_0}{\hbar}(c-u) = \frac{\partial}{\partial E}\frac{E}{c\hbar}\frac{1-u}{1-c} = \frac{\Delta k}{E} \end{cases} \tag{C3}$$

Substituting the results **(C3)** into derivative **(C2)** and then the final expression into **(C1)**, we get:

$$\frac{\partial arg(z)}{\partial E} = \frac{2m_0\Delta kE(1-2\chi^2)+2\hbar^2\chi^2\Delta k(2\Delta k^2-1)}{\hbar^2\chi E[(\Delta k^2+\chi^2)^2+4\Delta k^2\chi^2]} \tag{C4}$$

from which we obtain the particle phase time **(16)**.

## Appendix D

The argument function of the secondary component of spinor **(4)** is:

$$arg(z^{2°}) = arctg\left[-2\beta^2\frac{\chi\Delta k}{\Delta k^2+\chi^2}\right] \tag{D1}$$

Before calculating its derivative, we have to explain the relativistic factor $\beta$ as the function of the particle energy. We proceed as follows:

$$E = \frac{1}{\sqrt{1-\beta^2}}m_0c^2 \Rightarrow m_0^2c^4 = E^2(1-\beta^2) \Rightarrow \beta^2 = \frac{E^2-m_0^2c^4}{E^2} \tag{D2}$$

Now we calculate the derivative of function **(D1)** using the chain rule:

$$\frac{\partial arg(z^{2°})}{\partial E} = \frac{(\Delta k^2+\chi^2)^2}{(\Delta k^2+\chi^2)^2+4\beta^4\Delta k^2\chi^2}\frac{\partial}{\partial E}\left(-2\beta^2\frac{\chi\Delta k}{\Delta k^2+\chi^2}\right) =$$
$$= \frac{(\Delta k^2+\chi^2)^2}{(\Delta k^2+\chi^2)^2+4\beta^4\Delta k^2\chi^2}\left[\frac{\partial}{\partial E}\left(-2\frac{\chi\Delta k}{\Delta k^2+\chi^2}\right)+\frac{\partial}{\partial E}(\beta^2)\right] \quad \textbf{(D3)}$$

We have already calculated the first derivative in the square brackets in Appendix **C**. We calculate the second derivative using expression **(D2)**:

$$\frac{\partial}{\partial E}(\beta^2) = \frac{\partial}{\partial E}\frac{E^2-m_0^2 c^4}{E^2} = 2\frac{(1-\beta^2)}{E} \quad \textbf{(D4)}$$

Substituting the results **(D4)** and **(C2)** into **(D3)**, we get:

$$\frac{\partial arg(z^{2°})}{\partial E} = 2\frac{(\Delta k^2+\chi^2)^2}{(\Delta k^2+\chi^2)^2+4\beta^4\Delta k^2\chi^2}\frac{(1-\beta^2)}{E} + \frac{\partial arg(z)}{\partial E} \quad \textbf{(D5)}$$

from which we obtain the phase time **(17)** of the secondary spinor component.